\documentclass[preprint,12pt,numbers]{elsarticle}

\usepackage{mathtools}
\usepackage[utf8]{inputenc}
\usepackage{placeins}
\usepackage{multirow}
 \usepackage{pifont}
\usepackage{xcolor}
\usepackage{hyperref} 
\usepackage[all]{hypcap} 
\hypersetup{colorlinks=true,linkcolor=blue,citecolor=blue} 
\usepackage{graphicx}
\usepackage{enumitem}
\usepackage{afterpage}

\graphicspath{{figures/}}

\makeatletter
\def\ps@pprintTitle{%
  \let\@oddhead\@empty
  \let\@evenhead\@empty
  \let\@oddfoot\@empty
  \let\@evenfoot\@empty
}
\makeatother

\usepackage{nomencl}
\makenomenclature

\usepackage{etoolbox}
\usepackage{soul}
\usepackage{amsmath}
\renewcommand\nomgroup[1]{%
  \item[\bfseries
  \ifstrequal{#1}{A}{Acronyms}{%
  \ifstrequal{#1}{B}{Symbols}}
]}

\begin{document}

\pagenumbering{roman}

\nomenclature[B]{$\mu$}{dynamic viscosity}
\nomenclature[B]{$\theta$}{contact angle}
\nomenclature[B]{$\sigma$}{interfacial tension (IFT)}
\nomenclature[B]{$\sigma_{cut-oll}$}{cut-oll interfacial tension}
\nomenclature[B]{$r$}{pore throat radius}
\nomenclature[B]{$r_{mean}$}{mean pore throat radius}
\nomenclature[B]{$P$}{Capillary pressure}
\nomenclature[B]{$R^H$}{Hydraulic resistance}
\nomenclature[B]{$L$}{pore body length}
\nomenclature[B]{$Q$}{Flux}
\nomenclature[B]{$\triangle p$}{pressure drop}
\nomenclature[B]{$P_{in}$}{driving/inlet pressure}
\nomenclature[B]{$P_{out}$}{outlet pressure}
\nomenclature[B]{$V$}{velocity}
\nomenclature[B]{$\beta$}{tuning parameter}
\nomenclature[B]{$y,y0,A$}{fitting parameters}





\title{Hindered transport of spherical particles in cylindrical pores: The role of structural heterogeneity in rejection-permeability trade-offs}

\author[huji]{Debanik Bhattacharjee}
\author[technion]{Yaniv Edery}
\author[technion]{Guy Z. Ramon\corref{cor1}}
\ead{ramong@technion.ac.il}

\cortext[cor1]{Corresponding author}

\affiliation[huji]{
  organization={Edmond and Lily Safra Center for Brain Sciences, Hebrew University},
  city={Jerusalem},
  country={Israel}
}

\affiliation[technion]{
  organization={Department of Civil and Environmental Engineering, Technion -- Israel Institute of Technology},
  city={Haifa},
  postcode={32000},
  country={Israel}
}

\pagenumbering{arabic}
\begin{abstract}
Membrane separations rely on balancing rejection and permeability. Extensive work has clarified how pore structure and operating conditions control each quantity in idealized or weakly heterogeneous systems. However, it remains unclear how this trade-off emerges in strongly heterogeneous media, where coupled distributions of pore and particle sizes shape the local balance between advection and diffusion and generate substantial variability in performance among distribution realizations. Here we present a steric hindered-transport framework for spherical particles in cylindrical pores that explicitly resolves both single and coupled dual heterogeneity in size distributions. We show that the ensemble-averaged rejection increases with the particle-pore aspect ratio $\lambda$ and with the P\'eclet number $Pe$, while advection enhances steric exclusion by up to $\sim$20\% at intermediate $\lambda$. Dual heterogeneity broadens the distribution of effective $Pe$, increases the variability and incidence of anomalous rejection trends, while systematically shifting the rejection-permeability trade-off toward higher permeability at fixed rejection. These results suggest that controlled heterogeneity can serve as a design lever to expand the attainable operating space for simultaneous high selectivity and high throughput.

\end{abstract}
\maketitle

\section*{INTRODUCTION}

Membrane-based separations have become indispensable across diverse fields such as water
purification~\cite{shannon2008,werber2016,elimelech2011}, pharmaceutical and biotechnology
processing~\cite{charcosset2006}, food bioprocessing~\cite{charcosset2021}, and sustainable
resource recovery~\cite{hube2020,chang2020}. Their appeal lies in a unique combination of
scalability, modularity, and energy efficiency, which makes them preferable over thermally
driven or chemically intensive separation methods~\cite{shannon2008,werber2016}. As the
demand for clean water, biopharmaceuticals, and circular-material processes continues to grow,
understanding and improving membrane transport mechanisms remains a central scientific and
engineering challenge.

At the heart of membrane performance is the balance between contaminant rejection and
permeability~\cite{zhang2020,lee2011,lim2021}. Rejection determines the membrane’s selectivity,
while permeability dictates throughput and energy cost. Maximizing both simultaneously,
however, is nontrivial: membranes that achieve strong rejection often suffer from limited
flux, whereas highly permeable membranes may fail to exclude critical solutes. This canonical
trade-off is now well documented across pressure- and solution-diffusion-driven
systems~\cite{geise2011,zhang2016}, and has been analyzed for porous MF/UF membranes in
terms of permeability-selectivity relationships~\cite{Siddiqui2016,Mehta2005,Kanani2010,Mochizuki1993}.
This fundamental trade-off is governed by the physics of hindered transport, where steric interactions, transport regimes (diffusion versus convection), and membrane structural features collectively dictate solute-membrane interactions~\cite{deen1987,dechadilok2006}.

Classical hindered-transport theories have provided foundational insights into steric
exclusion in membrane pores~\cite{renkin1954,anderson1974,davidson1988,phillips1990,bungay1973}.
These models typically describe rejection in terms of the solute-to-pore size ratio
(aspect ratio). 
These hindered-transport concepts were later translated into membrane-separation settings, including ultrafiltration sieving analyses, pore-size-distribution inference from retention data, and predictive nanofiltration models for solute rejection and transport through confined pores \cite{Aimar1990,Mochizuki1992,BowenMukhtar1996,BowenMohammadHilal1997}. When the aspect ratio is much smaller than one, solutes behave as if
unhindered and pass through the pores freely; as the aspect ratio approaches unity,
steric exclusion becomes nearly complete. 
Transport dynamics are further influenced by the Péclet number, which characterizes the competition between convective and diffusive transport. Taken together, these parameters capture the classic relationship between rejection and permeability that underpins modern models of hindered transport. 
Recent work has begun to synthesize these ideas into broader structure-performance frameworks that link pore-scale morphology, transport physics, and macroscopic separation metrics~\cite{Mondal2019}.

Despite their elegance, many of these models rely on simplifying assumptions that rarely
hold in practice. A critical limitation is the assumption of uniformity: pores are often
idealized as monodisperse and cylindrical, and solutes as having a single well-defined size.
Real systems, however, exhibit substantial heterogeneity. 
Solutes span wide size distributions, from small proteins~\cite{habert2024} and ions~\cite{tang2020} to virus-scale biomacromolecules~\cite{yadav2022} and colloids~\cite{dickhout2017}, and pore structures vary because of fabrication routes~\cite{shiohara2021,gates1999}. 
This heterogeneity can fundamentally alter separation performance~\cite{Siddiqui2016,nakao1994,BowenWelfoot2002,gu2023,Mochizuki1993}.
In particular, permeability-selectivity analyses by Zydney and co-workers have shown how
pore-size distributions, pore geometry (cylindrical vs.\ slit-shaped), and membrane
stretching can shift trade-off curves and, in some cases, move membranes closer to putative performance frontiers~\cite{Mochizuki1993,Mehta2005,Kanani2010}.

Several studies have already shown that pore-size distributions can strongly influence membrane selectivity and transport. 
In ultrafiltration, retention data have long been used to infer membrane pore-size distributions, and theoretical analyses demonstrated that the breadth of the distribution affects asymptotic sieving, hindered solute diffusivity, and hydraulic permeability \cite{Aimar1990,Mochizuki1993,Zydney1994}. 
Related nanofiltration modeling studies likewise showed that the broadness of the pore-size distribution can substantially alter predicted rejection, often through the contribution of the large-pore tail \cite{BowenWelfoot2002}. More recent permeability-selectivity analyses further quantified how pore geometry and pore-size-distribution width reshape trade-off curves \cite{Mehta2005,Kanani2010,Siddiqui2016}. 
However, these studies mostly treat pore heterogeneity at the membrane level while assuming a single solute size, or they examine one source of heterogeneity at a time. The combined effect of simultaneous pore and particle size distributions on hindered-transport-governed rejection and on realization-to-realization variability remains much less directly resolved.

The more realistic scenario of dual heterogeneity, where both pores and particles follow
distributions rather than single values, remains comparatively underexplored, despite being ubiquitous in natural and engineered systems. Dual heterogeneity introduces new complexity:
systems with identical mean aspect ratios can yield markedly different rejection profiles depending on the breadth and overlap of particle and pore
distributions~\cite{Siddiqui2016,BowenWelfoot2002,gu2023}. Such variability can lead to anomalous transport behaviors (early breakthrough with long-time tailing, non-Gaussian residence times, and sub- or super-diffusive spreading), complicating predictions based on mean values alone ~\cite{dagan_bifurcating_2024,edery_origins_2014}. At the same time, if properly understood, heterogeneity can be leveraged as a design
advantage, offering routes to relax trade-off boundaries and design membranes optimized for selective yet high-throughput separations~\cite{Siddiqui2016,gu2023,Mondal2019}. 

In this work, we address this more specific gap by employing a steric hindered-transport model that resolves the coupled roles of aspect ratio, P\'eclet number, pore-size heterogeneity, and particle-size heterogeneity. Rather than revisiting pore-size-distribution effects alone, our focus is on how simultaneous dual heterogeneity reshapes rejection, the effective distribution of local transport conditions, and the rejection-permeability trade-off. 
We first revisit uniform systems to establish baselines consistent with classical
theory~\cite{deen1987,dechadilok2006}. 
We then introduce single heterogeneity
(heterogeneous particles with homogeneous pores) to quantify how individual sources of
variability alter ensemble-averaged rejection ($\chi$), mean aspect ratio ($\lambda$), and P\'eclet-number distributions. Building on this, we incorporate dual heterogeneity to capture realistic scenarios in which both pores and particles are heterogeneous. Finally, we evaluate how heterogeneity affects the rejection-permeability trade-off across targeted rejection windows, drawing explicit connections to permeability-selectivity frontiers discussed in prior membrane-structure studies~\cite{Siddiqui2016,Mochizuki1993,Mehta2005,Kanani2010,Mondal2019}.

Our results reveal that dual heterogeneity can, counterintuitively, elevate permeability
without degrading rejection, thus providing a mechanistic basis for viewing structural
variability as a potential design advantage rather than merely a defect.
Most of these studies assume log-normal pore-size distributions, which are consistent with many experimentally measured ultrafiltration membranes and capture the strongly right-skewed, heavy-tailed nature of such structures~\cite{Mochizuki1993,Kanani2010}. 
While physically grounded, log-normal distributions also couple distribution width to skewness and tail weight: as the distribution widens, a growing population of very large pores can dominate the hydraulic response. 
This coupling makes it difficult to separate the effect of overall heterogeneity
from the influence of a small number of extreme, high-flux pathways. 
In the present work, we adopt a complementary strategy aimed at mechanistic isolation rather than strict microstructure statistics, modeling both pore and particle radii using normal distributions with prescribed means and standard deviations, so that heterogeneity strength can be tuned without simultaneously changing skewness or tail weight. 
This choice enables controlled comparisons between single heterogeneity (particles heterogeneous, pores nearly uniform) and dual heterogeneity (both heterogeneous). 
It also allows us to quantify not only ensemble-averaged trends but also realization-to-realization variability at fixed rejection windows. 
In the next section, we detail the steric hindered-transport model, define the governing parameters (including $\lambda$ and $Pe$), and describe the procedure used to generate and analyze the heterogeneous ensembles.

\section{Model and Methodology}

\begin{figure*}
\centering
\includegraphics[width=1.0\linewidth]{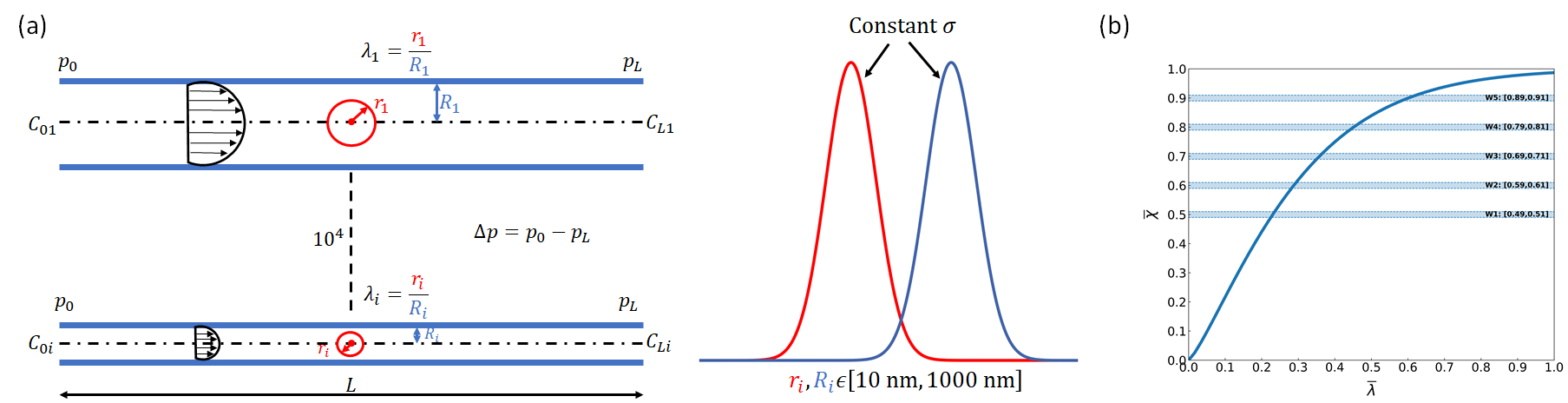}
\caption{(a) Schematic of single-pass transport of spherical particles through a cylindrical pore of length $L$ under a uniform pressure drop, $\Delta p = p_{0}-p_{L}$. For each configuration, $10^{4}$ particle-pore pairs are generated with radii $r_{i}$ and $R_{i}$ drawn independently from normal distributions with mean values in $[10,1000]~\text{nm}$ and with a common standard deviation $\sigma$, subject to $\lambda_i=r_i/R_i \le \lambda_{\max}<1$. Particles are advected by a Hagen-Poiseuille flow and constrained to the pore centerline between inlet $C_{0i}$ and outlet $C_{Li}$, and the configuration is summarized by its mean aspect ratio $\overline{\lambda}$. The Gaussian curves on the right illustrate the particle (red) and pore (blue) size distributions,
(b) On a representative ensemble-averaged rejection curve, $\overline{\chi}$ versus averaged aspect ratio, $\overline{\lambda}$, shaded horizontal bands (W1-W5) indicate the sampling windows used to compare permeability at matched rejection levels: $\overline{\chi}\in[0.49,0.51]$, $[0.59,0.61]$, $[0.69,0.71]$, $[0.79,0.81]$, and $[0.89,0.91]$.}

\label{fig:1} 
\end{figure*}

\subsection{Hindered transport model} 
\noindent 
We consider the transport of spherical particles in cylindrical pores, as illustrated in Fig.~\ref{fig:1}. 
The schematic highlights the upstream and downstream concentrations ($C_{0i}$ and $C_{Li}$), the particle radius $r_i$, the pore radius $R_i$, and the pore thickness $L$. The particle-pore aspect ratio, a central descriptor of steric hindrance, is defined as
\begin{equation}
\lambda_i = \frac{r_i}{R_i}.
\end{equation}
When $\lambda_i \ll 1$, particles are much smaller than the pores and traverse with minimal restriction. 
Conversely, as $\lambda_i \to 1$, the pore becomes effectively blocked for transport, leading to near-complete rejection. 
\\
At equilibrium, steric exclusion limits the concentration of particles within pores. 
For purely steric interactions, the equilibrium partition coefficient is 
\begin{equation}
\phi_i = (1-\lambda_i)^2,
\end{equation}
which corresponds to the fraction of the pore cross-section accessible to particle centers \cite{deen1987}. 
This quadratic form demonstrates how even moderate increases in $\lambda_i$ strongly reduce pore accessibility. 
\\
\\
Transport dynamics is governed by the balance between convection and diffusion, captured by the Péclet number:
\begin{equation}
Pe_i = \frac{vR_i}{D_i}=\frac{\Delta p R_i^2}{8 \eta D_i},
\end{equation}
where the Hagen-Poiseuille velocity, $v$ is driven by $\Delta p$, the applied pressure, $\eta$ is the solvent viscosity, $R_i$ and $D_i = \tfrac{k_\mathrm{B}T}{6 \pi \eta r_i}$ is the Stokes-Einstein diffusivity of a particle of radius $r_i$. 
Physically, small particles (large $D_i$) exhibit diffusion-dominated transport ($Pe_i \ll 1$), while large particles or high driving pressures correspond to a convection-dominated regime ($Pe_i \gg 1$). 
\\
\\
The resulting downstream-to-upstream concentration ratio is expressed as \cite{deen1987,dechadilok2006}

\begin{equation}
\frac{C_{Li}}{C_{0i}} = \frac{W_i}{1 - e^{-Pe_i} + W_i e^{-Pe_i}},
\end{equation}
where $W_i$ is the hindrance factor for convection. 
The corresponding rejection is then
\begin{equation}
\chi_i = 1 - \phi_i \left( \frac{W_i}{1 - e^{-Pe_i} + W_i e^{-Pe_i}} \right),
\end{equation}
showing that rejection depends simultaneously on steric partitioning ($\phi_i$), convective-diffusive competition ($Pe_i$), and the convective hindrance factor ($W_i$) \cite{deen1987,dechadilok2006}. 
\\
The convective hindrance factor is defined as
\begin{equation}
W_i = \frac{\phi_i (2-\phi_i)\,K_{si}}{2K_{ti}},
\end{equation}
where \(K_{ti}\) and \(K_{si}\) are hydrodynamic correction functions for the classical centerline hindered-transport problem of a rigid spherical particle in a cylindrical pore under laminar Hagen--Poiseuille flow with purely steric interactions. In this formulation, off-center migration, electrostatic and chemical interactions, and wall effects beyond these empirical correction factors are neglected \cite{bungay1973}. The correction functions are given by
\begin{align}
K_{ti} &=
\frac{9}{4}\pi^2\sqrt{2}\,(1-\lambda_i)^{-2.5}
\Big[1+a_1(1-\lambda_i^2)+a_2(1-\lambda_i^2)^2\Big]
+ a_3 + a_4\lambda_i + a_5\lambda_i^2 + a_6\lambda_i^3 + a_7\lambda_i^4,
\\
K_{si} &=
\frac{9}{4}\pi^2\sqrt{2}\,(1-\lambda_i)^{-2.5}
\Big[1+b_1(1-\lambda_i)+b_2(1-\lambda_i)^2\Big]
+ b_3 + b_4\lambda_i + b_5\lambda_i^2 + b_6\lambda_i^3 + b_7\lambda_i^4,
\end{align}
with numerical coefficients
\begin{equation}
\begin{aligned}
a_1 &= -\frac{73}{60}, &
a_2 &= \frac{77293}{50400}, &
a_3 &= -22.5083, &
a_4 &= -5.6117, \\
a_5 &= -0.3363, &
a_6 &= -1.216, &
a_7 &= 1.647, & & \\
b_1 &= -\frac{7}{60}, &
b_2 &= -\frac{2227}{50400}, &
b_3 &= 4.018, &
b_4 &= -3.9788, \\
b_5 &= -1.9215, &
b_6 &= 4.392, &
b_7 &= 5.006. & &
\end{aligned}
\end{equation}
The volumetric flow contribution of each pore is described by its hydraulic permeance, given via the Hagen-Poiseuille relation
\begin{equation}
P_i = \frac{\pi R_i^4}{8\eta LA_i},
\end{equation}
which shows that the area-normalized single-pore permeance scales as $R_i^2$.
In heterogeneous membranes, this scaling implies that a small fraction of larger pores can dominate the overall flux, as visually represented by the pore-size distribution in Fig.~\ref{fig:1}. 
In what follows, all lengths are non-dimensionalized by the pore length \(L\), which is set to unity in the simulations. The resulting transport coefficients should therefore be interpreted as permeability-like quantities. 
Because \(L\) appears only as a global prefactor, the relative trends and the comparison between single and dual heterogeneity are unaffected by the choice of physical membrane thickness. 
Permeance values for a membrane of thickness \(L_{\text{phys}}\) are recovered by dividing by \(L_{\text{phys}}\).

\paragraph*{Model assumptions and limitations.}
The framework forming the base of our analysis relies on the following simplifying assumptions: (i) particles are spherical and transported strictly along pore centerlines; (ii) interactions are purely steric, neglecting electrostatic or chemical effects; (iii) hydrodynamic interactions with pore walls are neglected beyond the empirical hindrance factors; (iv) velocity profiles are assumed to be parabolic and laminar, consistent with Hagen-Poiseuille flow; and (v) operation is single-pass (once-through), with no retentate recycle or staging. 
While these approximations enable analytically tractable treatment of hindered transport in cylindrical pores, they neglect effects such as off-center trajectories, wall-induced migration, and non-steric partitioning, as well as system-level phenomena pertinent to multi-pass or recirculating modules. 
Nevertheless, despite its known limitations, this model remains a standard framework for studying hindered transport. The present work advances this framework toward more realistic heterogeneous scenarios by analyzing the coupled roles of aspect ratio, Péclet number, and heterogeneity, thereby providing mechanistic insight into selective membrane design.

\subsection{Computational methodology}
\noindent
\noindent In the forthcoming analysis, each configuration comprises $N = 10^4$ independently sampled particle-pore pairs, specified by particle radius $r_i$ and pore radius $R_i$. 
Radii are drawn independently, from normal distributions with fixed standard deviation $\sigma$, while the means $(\bar r, \bar R)$ are varied systematically. 
We deliberately adopt normal, rather than log-normal, size distributions, which provides a minimal, symmetric representation of variability where the standard deviation, $\sigma$, acts as a direct measure of heterogeneity in the bulk population, and increasing $\sigma$ broadens the core of the distribution without generating a heavy right tail. 
In contrast, for log-normal pore-size distributions, modest changes in $\sigma$ strongly enhance the right tail and can cause the overall flux to be dominated by a small subset of very large pores, making it difficult to ascribe changes in membrane performance to the heterogeneity experienced by typical pores and particles~\cite{Mochizuki1993,Kanani2010}.

Two classes of heterogeneity are considered. 
In \textit{single heterogeneity}, particles vary in size according to their distribution ($\sigma_{\overline{r_i}} = 10\,\text{nm}$) while pores are essentially homogeneous ($\sigma_{\overline{R_i}} = 10^{-3}\,\text{nm}$). 
In \textit{dual heterogeneity}, both particles and pores vary with finite and equal standard deviations ($\sigma_{r_i} = \sigma_{R_i} = 10\text{nm}$), capturing the realistic situation of both structural and solute variability. 
We enforce a cutoff \(\lambda \le \lambda_{\max} < 1\) to exclude non-physical overlaps (\(\lambda>1\)). 
Although realizations with $\lambda>1$ are excluded by construction, this does not produce an observable skew in the final sampled set. As shown later, the mean-$\lambda$ histogram remains approximately equispaced. This is because realizations are repeatedly regenerated until comparable numbers of accepted samples are obtained across bins, thereby minimizing bias in the retained distribution.
\noindent
For each configuration $j$, the mean of any generic transport-related quantity, $q$, (such as rejection efficiency $\chi$ or aspect ratio $\lambda$) is computed for the ensemble as follows:
\begin{equation}
\overline{q}_j = \frac{1}{N} \sum_{i=1}^{N} q_{ij}, \qquad j = 1, \dots, 10^4, \;\; N=10^4,
\end{equation}
where $\overline{q}_j$ denotes the average over all particle-pore pairs in configuration $j$. 
The collection of $\{\overline{q}_j\}$ across all configurations yields an ensemble distribution of mean properties. 
This ensemble-based treatment highlights not only average values but also the variability across systems, thereby connecting microscopic pair-level behavior with macroscopic, configuration-level performance.  

\noindent
In order to establish the impact of pore heterogeneity on selectivity, while providing a fair comparison of permeability across systems, we evaluate the permeability at identical rejection.
This is achieved by defining a narrow sampling windows on the \(\overline{\chi}\)-\(\overline{\lambda}\) plane
(Fig.~\ref{fig:1}c), such that each horizontal band fixes \(\overline{\chi}\) (rejection) while allowing variation in \(\overline{\lambda}\) and pore geometry.
Five representative windows are used:
\[
\mathrm{W1}:[0.49,0.51],\quad
\mathrm{W2}:[0.59,0.61],\quad
\mathrm{W3}:[0.69,0.71],\quad
\mathrm{W4}:[0.79,0.81],\quad
\mathrm{W5}:[0.89,0.91].
\]

\noindent
Configurations that fall within a given window $W$ are identified, and the cumulative permeance contributed by all pores in that subset is calculated as
\begin{equation}
\sum_{i} P_i = \sum_{i \in W} P_i, \qquad W \in \{\mathrm{W1},\mathrm{W2},\mathrm{W3},\mathrm{W4},\mathrm{W5}\}.
\end{equation}
This procedure allows for a systematic comparison of how heterogeneity influences permeance at equivalent rejection levels, thereby disentangling the effects of particle variability, pore variability, and their combined influence.

\FloatBarrier
\section{Results and Discussion}

The theoretical framework outlined in the previous section provides a quantitative basis for analyzing the impact of particle- and pore-size heterogeneity on hindered transport of spherical particles through cylindrical pores. 
In this section, we use this model to examine how rejection efficiency and permeability depend on the particle-pore aspect ratio, the Péclet number, and the degree of structural heterogeneity. 
We first analyze the behavior of heterogeneous spherical particles in a homogeneous array of identical pores, which serves as a reference case. 
Pore-scale heterogeneity is then introduced to assess its influence on transport characteristics. 
Finally, we examine the rejection-permeability trade-off, a central determinant of membrane performance. 
Unless stated otherwise, lengths are non-dimensionalized by the pore length, so that the dimensionless pore length is unity. Accordingly, the transport coefficients reported by this framework are thickness-independent permeability-like quantities. For a membrane with physical thickness $L_{\text{phys}}$, the corresponding permeance is obtained by dividing by $L_{\text{phys}}$.

\begin{figure*}[!htbp]
\centering
\includegraphics[width=0.75\linewidth]{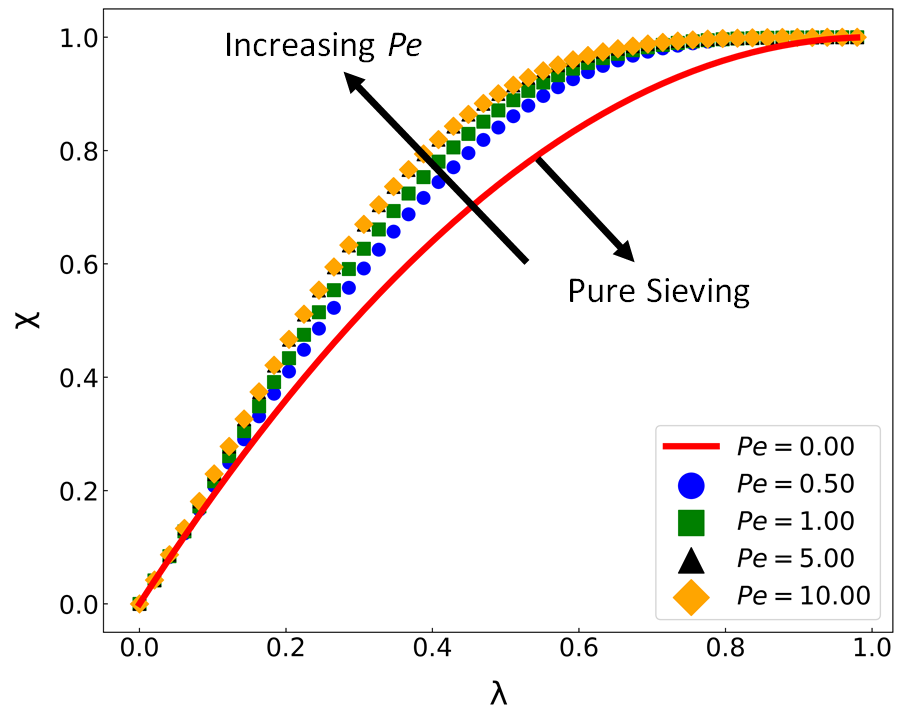}
\caption{Rejection ($\chi$) as a function of the aspect ratio ($\lambda = r/R$) for single-pass transport of spherical particles through homogeneous cylindrical pores at different Péclet numbers ($Pe$), spanning diffusion- to convection-dominated regimes. 
Here, $\lambda$ is prescribed directly as the ratio of a single representative particle radius $r$ to pore radius $R$ (i.e., a monodisperse suspension and a given membrane pore size), and no size distributions are considered. The solid red line denotes the sieving limit ($Pe = 0$), while the symbols show the impact of increasing $Pe$ ($0.5 \le Pe \le 10$).  Hindered convection has the strongest impact in the intermediate range $0.2 \leq \lambda \leq 0.6$, where $\chi$ is enhanced by up to $\approx 15$-$20\%$. }

\label{fig:2} 
\end{figure*}

Fig.~\ref{fig:2} presents the rejection ($\chi$) as a function of the aspect ratio ($\lambda = r/R$) for spherical particles in cylindrical pores under different Péclet numbers ($Pe$). 
In this monodisperse baseline calculation, $\lambda$ is prescribed directly as the ratio of a single particle radius, $r$, to a single pore radius, $R$, (i.e. $r/R=\lambda$), and no particle or pore size distributions are included. 
Across all examined $Pe$, $\chi$ increases monotonically with $\lambda$, approaching zero as $\lambda \to 0$ and unity as $\lambda \to 1$. 
This behavior reflects the steric exclusion imposed by the pore geometry: particles much smaller than the pore ($\lambda \leq 0.1$) pass essentially unhindered, whereas particles comparable to the pore size ($\lambda \geq 0.8$) are almost completely rejected.
Qualitatively, this trend is similar to that observed in the case of pure sieving. 

\begin{figure*}[t]
    \centering
    \includegraphics[width=0.8\linewidth]{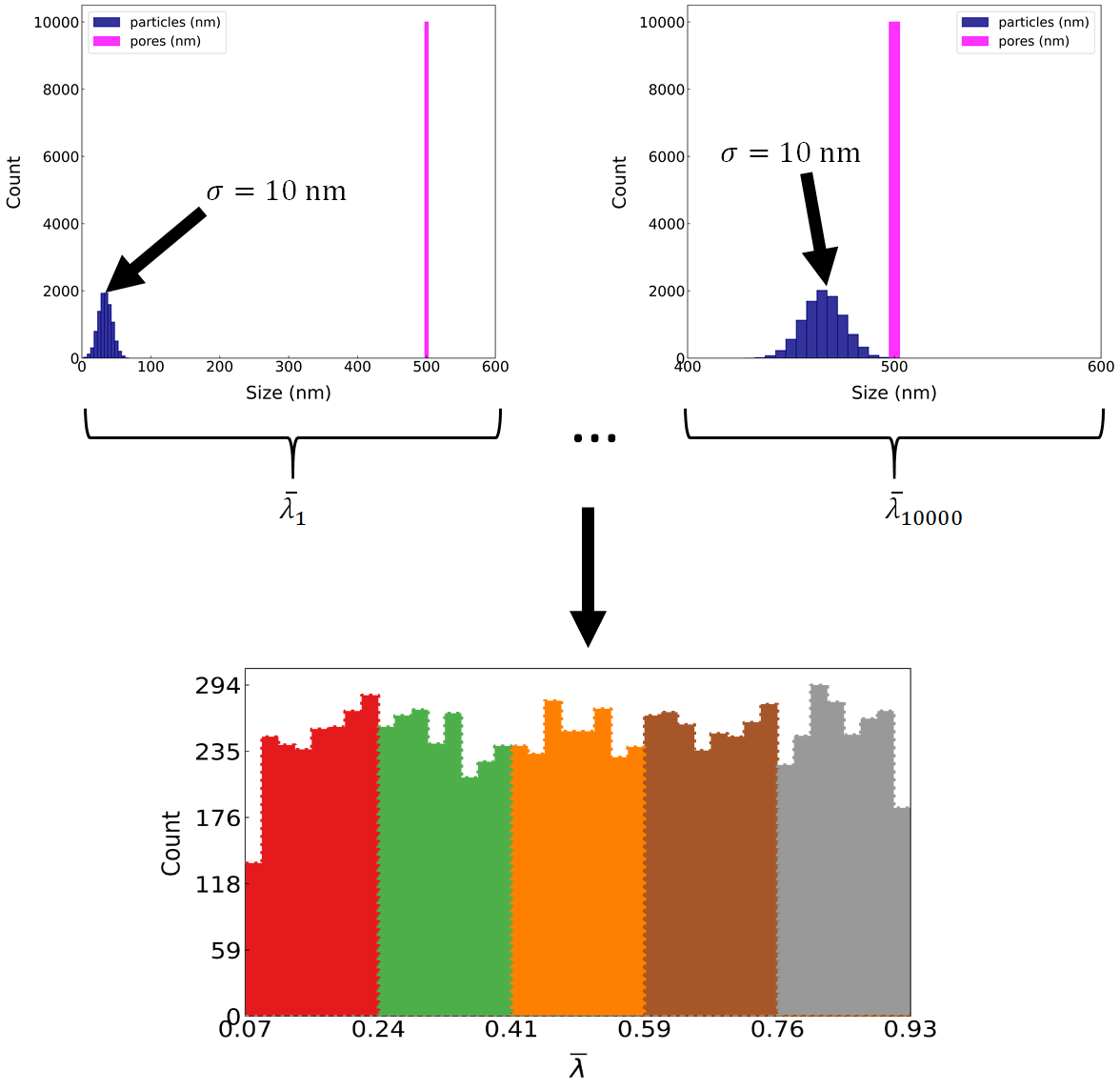}
    \caption{\textbf{Single heterogeneity ensemble construction.}
Top: two representative particle-pore size distributions drawn from the $10^{4}$ independently sampled configurations. In each configuration, particle radii are heterogeneous, $r \sim \mathcal{N}(\bar r,\sigma_r)$ with $\sigma_r=10~\mathrm{nm}$ and $\bar r$ scanned from $10$ to $1000~\mathrm{nm}$ (navy histogram), while pore radii are effectively monodisperse, $R \sim \mathcal{N}(500~\mathrm{nm},0~\mathrm{nm})$ (magenta spike).
Bottom: histogram of $\bar{\lambda}$ over all $10^{4}$ configurations; colored bands mark the $\bar{\lambda}$ intervals used to group the ensemble in subsequent panels.}
    \label{fig:3} 
\end{figure*}

However, our model incorporates the effect of internal transport hindrance, highlighting the influence of $Pe$, which is most pronounced in the intermediate range $0.2 \leq \lambda \leq 0.6$, where convection and diffusion contribute comparably to transport. 
At $\lambda \simeq 0.3$, for example, $\chi$ increases from about $0.55$ for purely diffusive transport ($Pe = 0$) to nearly $0.70$ for $Pe = 10$. 
Similarly, at $\lambda \simeq 0.5$, rejection increases from $\chi \simeq 0.78$ ($Pe = 0$) to $\chi \simeq 0.95$ ($Pe = 10$). 
In this intermediate regime, convection increases particle retention by approximately 15--20\% compared with the pure-sieving limit.
For $\lambda \leq 0.1$, $\chi \leq 0.2$ across all $Pe$, while for $\lambda \geq 0.8$, $\chi \geq 0.95$, indicating that hydrodynamic effects are negligible compared to steric constraints in these limits. 
Moreover, the curves for $Pe \ge 5$ nearly collapse onto each other, signaling the approach to a convection-dominated regime in which further increases in $Pe$ is of little additional effect. 
This single-pore response provides the reference behavior against which the population-level effects of particle and pore size heterogeneity are interpreted in the following sections.

While Fig.~\ref{fig:2} illustrates the dependence of rejection on $Pe$ and $\lambda$ for an ideal membrane with homogeneous pores, real membranes are seldom of uniform pore-size. 
Variations in particle sizes, pore sizes, or both can strongly modify transport by locally changing both steric exclusion and the relative contribution of advection and diffusion. 
To capture these effects, we next analyze rejection under controlled single- and dual-heterogeneity conditions.
Fig.~\ref{fig:3} shows the construction of the single-heterogeneity ensemble. 
Here, particle size is varied while the pore radius is fixed, which produces a broad distribution of configuration-averaged aspect ratios $\bar{\lambda}$. 
We then partition the ensemble into the $\bar{\lambda}$ intervals indicated by the colored bands; these bins are used throughout the subsequent analysis to compare transport statistics across different aspect ratios.

\begin{figure*}[t]
    \centering
    \includegraphics[width=1.0\linewidth]{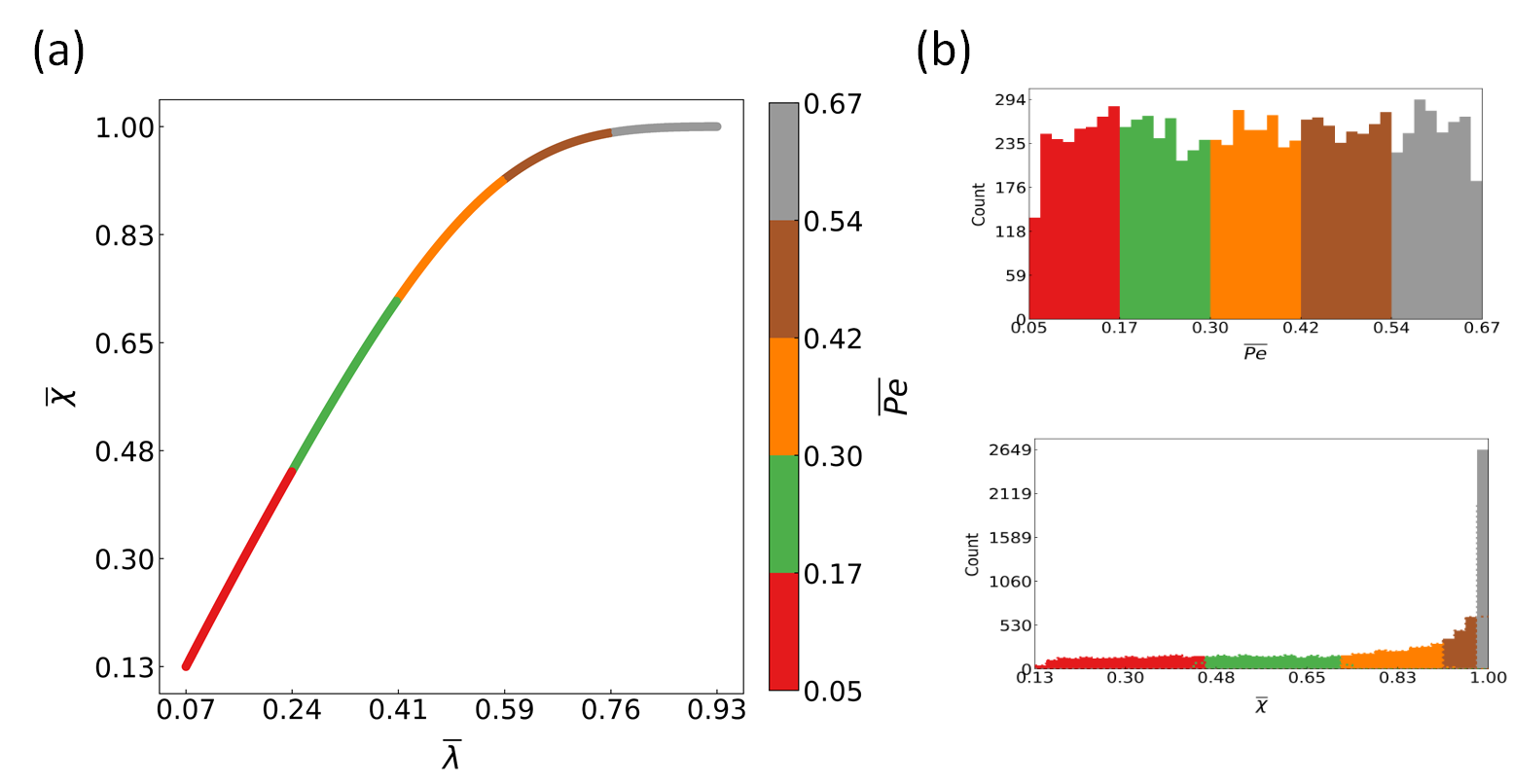}
    \caption{\textbf{Single-heterogeneity transport response.}
(a) Configuration-averaged rejection $\bar{\chi}$ versus configuration-averaged aspect ratio $\bar{\lambda}$ across $10^{4}$ independently sampled particle-pore configurations, (b) Ensemble histograms of $\overline{Pe}$ (top) and $\bar{\chi}$ (bottom); bars are colored by the same $\overline{Pe}$ intervals to connect transport regime to rejection outcome.
Particles are polydisperse with $r \sim \mathcal{N}(\bar r,\sigma_r)$, where $\bar r$ is scanned from $10$ to $1000~\mathrm{nm}$ and $\sigma_r=10~\mathrm{nm}$, while pores are effectively monodisperse with $R \sim \mathcal{N}(500~\mathrm{nm},0~\mathrm{nm})$.
Parameters: $\Delta p = 10^{-2}~\mathrm{Pa}$, $L=1$.}
    \label{fig:4} 
\end{figure*}

We now turn to quantify how these aspect-ratio bins translate into transport outcomes by computing, for each sampled configuration, the averaged rejection $\bar{\chi}$ and the associated mean P\'eclet number $\overline{Pe}$ (Fig.~\ref{fig:4}). As shown in Fig.~\ref{fig:4}a, $\bar{\chi}$ increases monotonically with $\bar{\lambda}$: for $\bar{\lambda}\ll 1$ particles access most of the pore cross-section, experience weak steric exclusion, and are largely transmitted (e.g.\ $\bar{\lambda}\leq 0.1 \Rightarrow \bar{\chi}\leq 0.2$). 
As $\bar{\lambda}$ approaches unity, the accessible cross-sectional area shrinks and particles become increasingly confined; steric exclusion dominates and rejection becomes nearly complete (e.g.\ $\bar{\lambda}\geq 0.8 \Rightarrow \bar{\chi}\geq 0.95$). 
Because the pore radii are effectively fixed, the mean velocity varies little across realizations; the modest spread in $\overline{Pe}$ (Fig.~\ref{fig:4}b, top) therefore mainly reflects changes in particle diffusivity with size. 
This yields a tight collapse onto a single $\bar{\chi}(\bar{\lambda})$ curve with minimal scatter. 
Fig.~\ref{fig:4}b shows how this geometry-to-function mapping shapes the ensemble statistics: while $\overline{Pe}$ remains confined to low values, the distribution of $\bar{\chi}$ is strongly skewed toward high rejection, reflecting the nonlinear, saturating dependence of $\bar{\chi}$ on $\bar{\lambda}$ that compresses a wide range of large-$\bar{\lambda}$ configurations into $\bar{\chi}\approx 1$.

\begin{figure*}[t]
    \centering
    \includegraphics[width=0.8\linewidth]{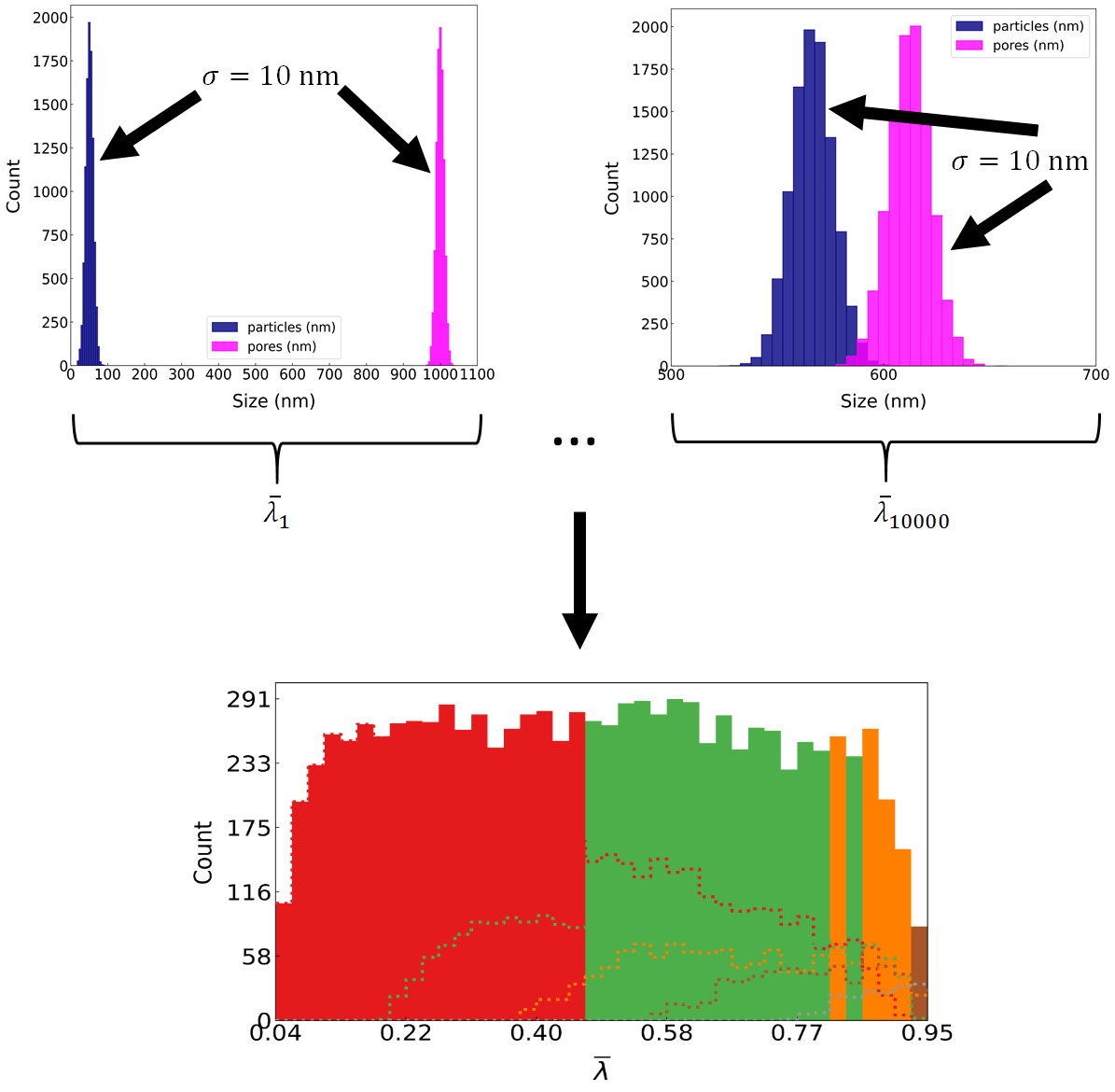}
    \caption{\textbf{Dual heterogeneity ensemble construction:}
Top: two representative particle-pore size distributions drawn from the $10^{4}$ independently sampled configurations (labels “1” and “10000” denote two illustrative indices from the full ensemble). In each configuration, both particle radii and pore radii are heterogeneous: $r \sim \mathcal{N}(\bar r,\sigma_r)$ (navy) and $R \sim \mathcal{N}(\bar R,\sigma_R)$ (magenta), with $\sigma_r=\sigma_R=10~\mathrm{nm}$; the means $\bar r$ and $\bar R$ are scanned across the ensemble. Braces indicate the corresponding configuration-averaged aspect ratio $\bar{\lambda}$ for each example.
Bottom: histogram of $\bar{\lambda}$ over all $10^{4}$ configurations; colored bands mark the $\bar{\lambda}$ intervals used to bin the ensemble in subsequent analyses.}
    \label{fig:5}
\end{figure*}

\begin{figure*}[t]
    \centering
    \includegraphics[width=1.0\linewidth]{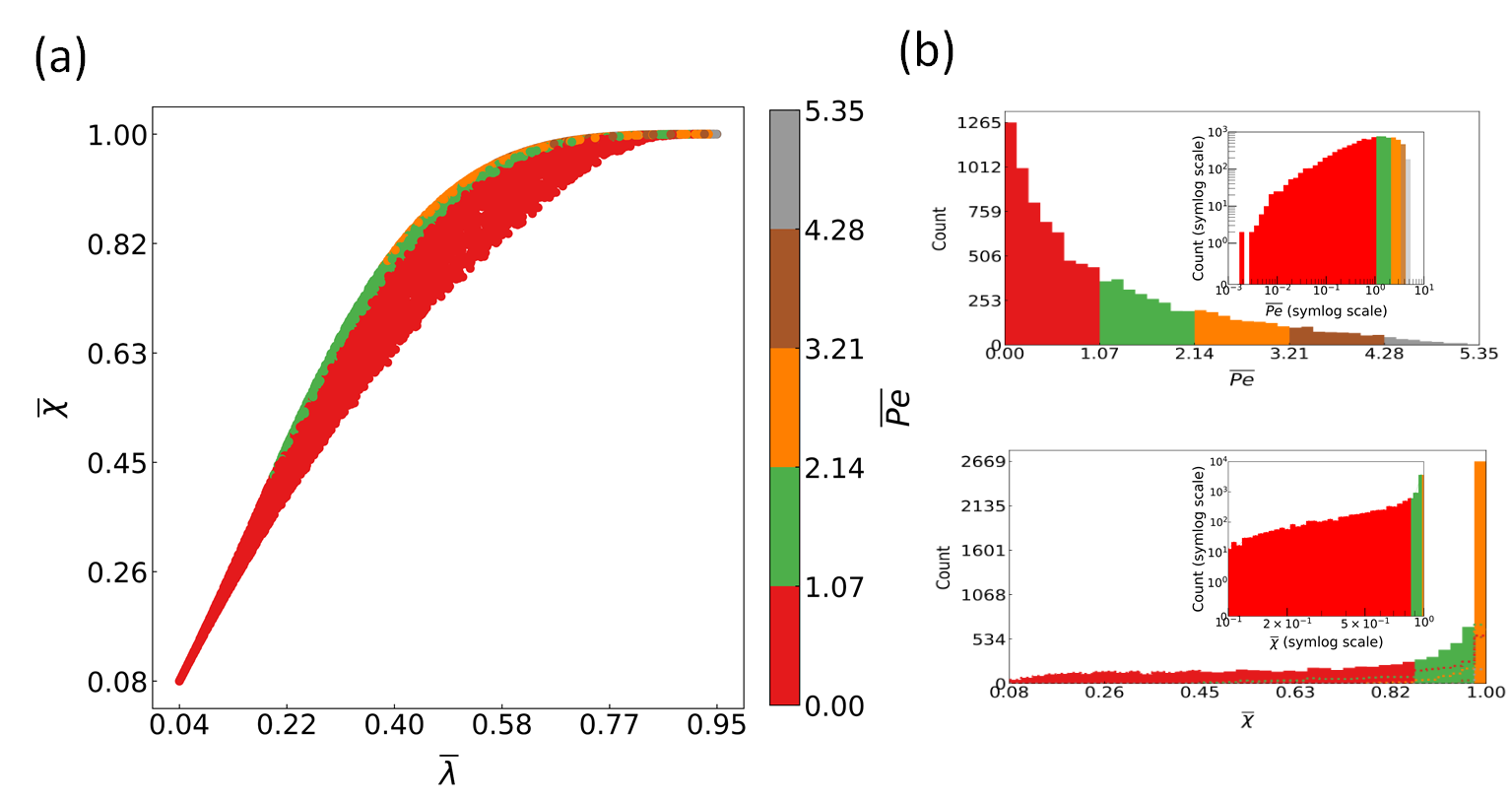}
    \caption{\textbf{Dual-heterogeneity transport response.}
(a) Configuration-averaged rejection $\bar{\chi}$ as a function of the configuration-averaged aspect ratio $\bar{\lambda}$ for $10^{4}$ independently sampled particle-pore configurations. Each point is colored by the corresponding configuration-averaged P\'eclet number $\overline{Pe}$, with $0 \leq \overline{Pe} \leq 5.35$. 
(b) Ensemble distributions of $\overline{Pe}$ (top) and $\bar{\chi}$ (bottom). Histogram bins are colored using the same $\overline{Pe}$ intervals as in panel (a), showing how transport regime maps onto the rejection statistics. The inset panels show the same distributions using symmetric logarithmic scaling, allowing both the high-count core and low-probability tails to be resolved on the same axes. The $\bar{\chi}$ distribution exhibits a clear tail toward high rejection values, consistent with the anomalous transport trends.
In this dual-heterogeneity ensemble, both particle and pore radii are polydisperse, with $r \sim \mathcal{N}(\bar{r},\sigma_r)$ and $R \sim \mathcal{N}(\bar{R},\sigma_R)$, where $\sigma_r=\sigma_R=10~\mathrm{nm}$, while the means $\bar{r}$ and $\bar{R}$ are scanned across the ensemble. Parameters: $\Delta p = 10^{-2}~\mathrm{Pa}$ and $L=1$.}
    \label{fig:6}
\end{figure*}

In the dual-heterogeneity ensemble (Fig.~\ref{fig:5}), both the particle and the pore sizes vary between realizations. 
This makes the hydraulic resistance configuration-dependent, unlike the single-heterogeneity case, where the pore geometry is fixed. 
For a given imposed pressure drop, the mean axial velocity in a cylindrical pore scales as $v \propto R^{2}$. 
Wider pores therefore carry faster flow and yield larger $\overline{Pe}$, while narrower pores yield smaller $\overline{Pe}$. Particle size also matters, as it changes the Stokes-Einstein diffusivity, $D\propto 1/r$, and sets the confinement through $\lambda=r/R$. 
A simple scaling then gives $Pe \sim vL/D \propto R^{2}r$, so variability in both $R$ and $r$ produces a broad spread in $\overline{Pe}$ (Fig.~\ref{fig:6}b, top), spanning diffusion-dominated to more advection-influenced regimes.

This broader sampling of transport conditions increases the dispersion in the calculated rejection. 
In Fig.~\ref{fig:6}a, $\bar{\chi}$ still increases monotonically with $\bar{\lambda}$, but the collapse onto a single curve is lost at intermediate aspect ratios. 
In this range, similar $\bar{\lambda}$ can correspond to very different $\overline{Pe}$ classes and therefore different rejection outcomes. 
For example, at $\bar{\lambda}\simeq 0.3$, $\bar{\chi}$ spans roughly $0.45$-$0.70$ across the ensemble. At $\bar{\lambda}\simeq 0.5$, the spread is larger, with some configurations reaching near-complete rejection ($\bar{\chi}\approx 0.95$) while others remain closer to $\sim 0.75$.

The ensemble histograms reinforce the decoupling between geometry and transport response: the same or similar values of the mean size ratio, $\bar{\lambda}$, can lead to different rejection outcomes, $\bar{\chi}$, because advection and diffusion are controlled by different heterogeneous variables. 
The distribution of $\overline{Pe}$ has a pronounced long tail (Fig.~\ref{fig:6}b, top), reflecting the combined variability in flow speed, set mainly by $R$, and diffusivity, set mainly by $r$. 
The $\bar{\chi}$ histogram is strongly skewed toward high rejection and exhibits a pronounced high-rejection tail, with contributions from multiple $\overline{Pe}$ classes (Fig.~\ref{fig:6}b, bottom). 
This tailing behavior shows that a subset of heterogeneous particle-pore configurations can produce disproportionately large rejection, even when the ensemble-level trend appears smooth. 
Thus, in the dual-heterogeneity setting, $\bar{\lambda}$ alone does not uniquely predict rejection. 
The functional response is set by the interplay between pore-controlled advection and particle-controlled diffusion.

\begin{figure*}[t]
    \centering
    \includegraphics[width=1.0\linewidth]{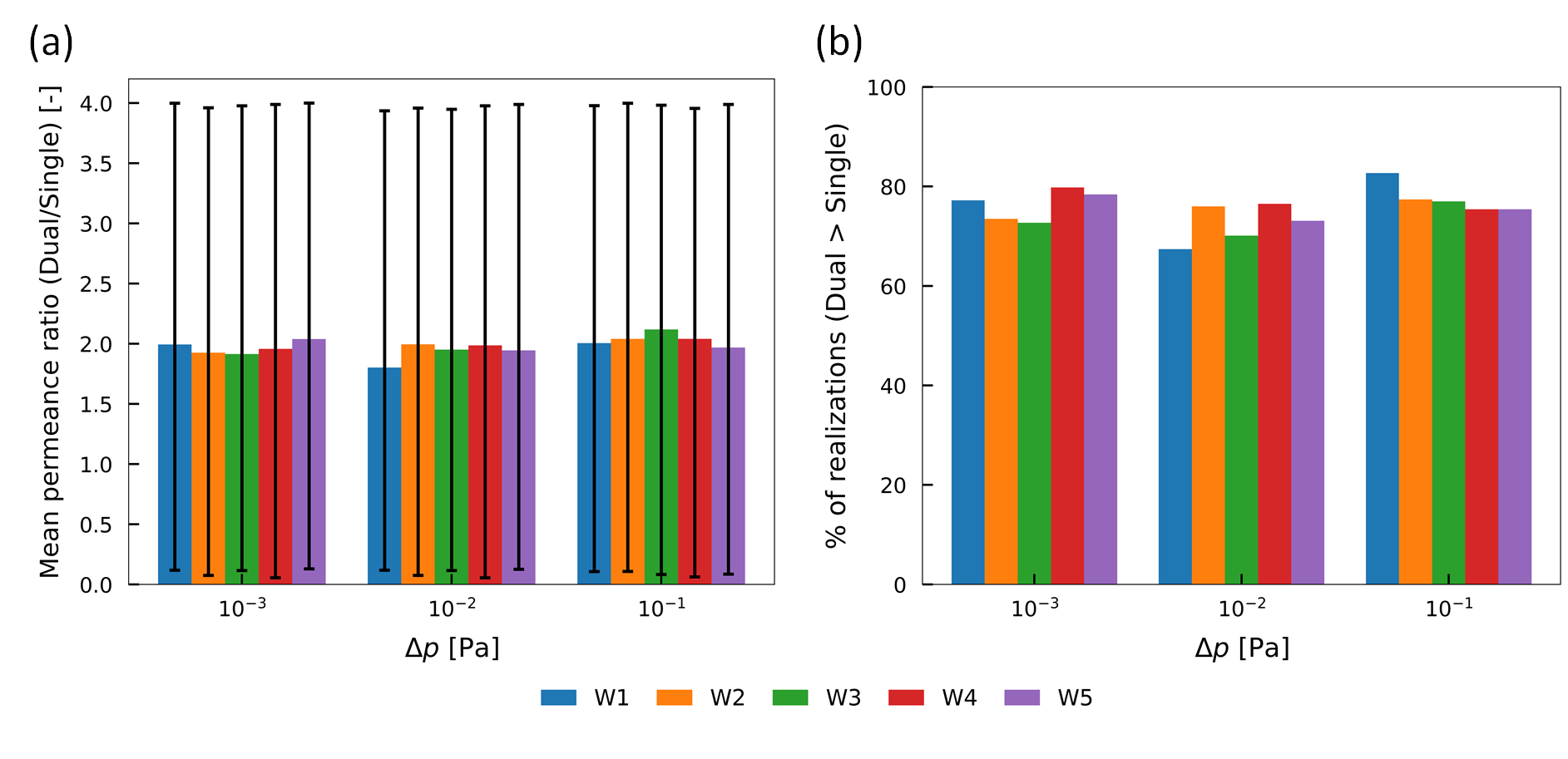}
    \caption{\textbf{Effect of dual heterogeneity on permeability at matched rejection levels.} (a) Mean permeability ratio, $\overline{\left(\frac{\sum_i P_i^{\mathrm{dual}}}{\sum_i P_i^{\mathrm{single}}}\right)}$, plotted versus applied pressure drop $\Delta p=10^{-3},\,10^{-2},\,10^{-1}\ \mathrm{Pa}$ for five narrow rejection windows defined by the mean rejection $\overline{\chi}$: W1 $[0.49,0.51]$, W2 $[0.59,0.61]$, W3 $[0.69,0.71]$, W4 $[0.79,0.81]$, and W5 $[0.89,0.91]$ (colors). “Single” heterogeneity denotes heterogeneous particles with nearly monodisperse pores ($\sigma_r=10\,\mathrm{nm}$, $\sigma_R=0\,\mathrm{nm}$), while “dual” heterogeneity includes variability in both particles and pores ($\sigma_r=\sigma_R=10\,\mathrm{nm}$). Bars show the mean across all realizations within each window, and error bars indicate the spread across realizations. Ratios exceeding unity across all windows and pressures indicate that adding pore-size variability typically boosts permeability at fixed rejection (often by about a factor of two). (b) Corresponding fraction of realizations (in \%) for which $\left(\sum_i P_i\right)_{\mathrm{dual}} > \left(\sum_i P_i\right)_{\mathrm{single}}$. Dual heterogeneity outperforms single heterogeneity in the majority of cases (roughly $\sim$70-83\% across the conditions shown), demonstrating a systematic shift of the rejection-permeability trade-off toward higher permeability.}
 
    \label{fig:7}
\end{figure*}

While Figs.~\ref{fig:4} and \ref{fig:6} highlight how heterogeneity shapes rejection, overall separation performance also depends on throughput. 
Real membranes must, ideally, deliver high rejection and high permeability at the same time. 
To probe this trade-off at fixed rejection, Fig.~\ref{fig:7} compares single and dual heterogeneity at matched mean rejection. We enforce this matching using five narrow windows, defined with reference to the $\bar{\chi}$ ranges shown in Fig.~\ref{fig:1}: $W_1=[0.49,0.51]$, $W_2=[0.59,0.61]$, $W_3=[0.69,0.71]$, $W_4=[0.79,0.81]$, and $W_5=[0.89,0.91]$.

A key point is how $\Delta p$ enters this comparison. 
The permeability used here is the \emph{hydraulic} permeability defined at the pore level in Eq.~(10) and aggregated as $\sum_i P_i$ (Eq.~(12)). 
For any fixed realization, this quantity is independent of $\Delta p$. 
The dependence on $\Delta p$ in Fig.~\ref{fig:7} arises because the \emph{conditioning on rejection} is $\Delta p$-dependent: for each operating pressure drop $\Delta p$, we first compute $\bar{\chi}$ for every realization (since $\bar{\chi}$ depends on $Pe$ and hence on $\Delta p$), then select only those realizations whose $\bar{\chi}$ falls within a given window $W_k$, and finally evaluate $\sum_i P_i$ for that selected subset. 
Thus, Fig.~\ref{fig:7} reports a \emph{conditional} permeability comparison,
\[
 \Biggl( \frac{\sum_i P_i^{\mathrm{dual}}}{\sum_i P_i^{\mathrm{single}}} \Biggr)_{\bar{\chi}\in W_k,\ \Delta p},
\]
rather than a direct $\Delta p$-dependence of permeability for a fixed geometry.

Figure~\ref{fig:7}a quantifies the impact of adding pore-size variability on permeability. It reports the mean ratio $\overline{(P_{\mathrm{dual}}/P_{\mathrm{single}})}$ for $\Delta p=10^{-3},10^{-2},10^{-1}$~Pa within each rejection window. The ratio exceeds unity for every window and every $\Delta p$. It also stays close to a twofold gain, $\overline{(P_{\mathrm{dual}}/P_{\mathrm{single}})}\simeq 1.8$-$2.1$. At $\Delta p=10^{-3}$~Pa, all windows sit near $\sim 2$. At $\Delta p=10^{-2}$~Pa, W1 is the lowest case (about $\sim 1.8$) while W2-W5 remain near $\sim 2$. 
At $\Delta p=10^{-1}$~Pa, the ratios again cluster near $\sim 2$, with a slight increase for intermediate-to-high rejection windows (e.g.\ W3-W4 reaching $\geq 2$).

The vertical bars in Fig.~\ref{fig:7}a show the \emph{minimum-maximum} range of $P_{\mathrm{dual}}/P_{\mathrm{single}}$ within each rejection window. 
They do not represent statistical uncertainty. 
The wide ranges indicate strong realization-to-realization variability even though $\bar{\chi}$ is tightly constrained by the window definition. 
Some realizations yield $P_{\mathrm{dual}}/P_{\mathrm{single}}\approx 1$, so dual heterogeneity offers little extra throughput at that rejection level. 
Others yield ratios well above the mean, so that dual heterogeneity enables substantially higher permeability, while still satisfying the same mean-rejection constraint. 
This spread is expected because dual heterogeneity enlarges the space of pore-particle pairings across realizations. As a result, different realizations can organize flux through different sets of more or less favorable pathways under the same global driving conditions, thereby broadening the range of transport outcomes. This wider minimum-maximum rejection window is therefore associated with the added variability introduced by the simultaneous presence of pore-size and particle-size distributions.

Figure~\ref{fig:7}b provides a complementary measure of how consistently dual heterogeneity improves throughput, showing the percentage of realizations for which $P_{\mathrm{dual}} > P_{\mathrm{single}}$. 
For every rejection window and every operating $\Delta p$, this percentage exceeds $50\%$. 
Overall, it falls in the range $\sim 67\%$ to $\sim 83\%$. 
At $\Delta p=10^{-3}$~Pa, dual heterogeneity produces higher permeability in roughly $\sim 73$-$80\%$ of realizations. 
At $\Delta p=10^{-2}$~Pa, W1 is the lowest case at about $\sim 67\%$, while W2-W5 typically lie around $\sim 70$-$77\%$. 
Finally, at $\Delta p=10^{-1}$~Pa, the percentages increase again, with W1 reaching $\sim 83\%$ and the remaining windows clustering near $\sim 75$-$78\%$.

Taken together, Fig.~\ref{fig:7}a-b convey a clear message. 
When $\bar{\chi}$ is held fixed, dual heterogeneity systematically shifts the system toward higher permeability. 
This is evident from the conditional mean ratio $\overline{(P_{\mathrm{dual}}/P_{\mathrm{single}})}$, which stays above unity across all windows and operating pressure drops. 
It is also evident from the majority fraction of realizations with $P_{\mathrm{dual}} > P_{\mathrm{single}}$ in every case.

\FloatBarrier

\section{Conclusions}

This study systematically examined the impact of structural heterogeneities on the hindered transport of spherical particles through cylindrical pores. 
For single heterogeneity (heterogeneous particle but homogeneous pore distributions), rejection ($\chi$) increased monotonically with the aspect ratio ($\lambda$), with convective hinderance enhancing exclusion in the intermediate regime ($0.2 \leq \lambda \leq 0.6$) by up to 20$\%$, while steric constraints dominate at both small and large $\lambda$. 
In contrast, under dual heterogeneity (heterogeneous particle and heterogeneous pore distributions), transport behavior was fundamentally altered. 
Whereas single heterogeneity yielded narrow Péclet number ($\overline{Pe}$) distributions and modest variability in rejection, dual heterogeneity extended the $\overline{Pe}$ range by more than an order of magnitude and produced a much broader spectrum of rejection outcomes, such that the same mean aspect ratio ($\overline{\lambda}$) could correspond to widely different rejection efficiencies. 

Beyond rejection alone, permeability analysis revealed that dual heterogeneity consistently outperformed single heterogeneity, at a given rejection. 
For fixed rejection windows, dual heterogeneity delivered higher permeability ($\sum_i P_i$) in more than 70-80$\%$ of cases across a wide range of applied pressures. 
This demonstrates that heterogeneity not only introduces greater variability, but also systematically shifts the rejection-permeability trade-off toward more favorable operating conditions. 

Overall, the results highlight that pore heterogeneity can provide a functional advantage by broadening the accessible operating space. 
By enabling simultaneous gains in both rejection and permeability, dual heterogeneity emerges as a promising design principle for improved membranes, where structural variability is deliberately harnessed rather than minimized. 
Future work should extend this framework to incorporate off-center particle trajectories and hydrodynamic interactions with pore walls, which were neglected under the present centerline-only assumption. 
Beyond single-pass transport, it will also be important to investigate the influence of heterogeneity on dynamic processes such as fouling, clogging, and concentration polarization, where variability in pore structure could delay critical transitions and improve long-term stability. 
Moreover, considering non-spherical or deformable particles, asymmetric or correlated size distributions, and spatially graded pore architectures could reveal new mechanisms by which structural disorder enhances selectivity and throughput. Finally, coupling simulations with experimental validation in realistic membrane materials will be essential for establishing heterogeneity as a robust design paradigm for membrane separations. 
Our use of normal size distributions is not intended to reproduce the detailed
pore-size statistics of any specific membrane, but rather to provide a symmetric baseline in which the role of distribution width can be examined in isolation. 
Many UF membranes indeed exhibit approximately log-normal pore-size distributions~\cite{Mochizuki1993,Kanani2010}, and we expect the qualitative features of the dual-heterogeneity trends reported here to extend to other unimodal distributions. 
Exploring how the shape of the pore and particle size distributions (e.g. log-normal versus normal) interacts with dual heterogeneity is an interesting direction for future work. In particular, such an analysis may help clarify the origin of the unresolved high-rejection tail observed in the $\bar{\chi}$ distribution, by determining whether rare but extreme rejection events arise primarily from the distributional tails of pore sizes, particle sizes, or their combinations.

\clearpage
\bibliographystyle{elsarticle-num}
\bibliography{Ref}

\end{document}